\input{aipcheck}

\documentclass[
    ,final            
  ]
  {aipproc}

\layoutstyle{8x11double}

\setcitestyle{numbers}
\usepackage{amsmath,amssymb,subfigure,wrapfig,verbatim} 
\usepackage{longtable}
\newcommand{\beq}{\begin{equation}}
\newcommand{\eeq}{\end{equation}}
\newcommand{\ee}{\epsilon}

\newcommand{\bea}{\begin{eqnarray}}
\newcommand{\eea}{\end{eqnarray}}
\newcommand{\ba}{\begin{array}}
\newcommand{\ea}{\end{array}}

\newcommand{\Tr}{{\rm Tr\,}}

\newcommand{\cA}{{\cal A}}

\newcommand{\hk}{\hspace{0.1cm}}

\newcommand{\rk}{\right)}
\newcommand{\lk}{\left(}

\newcommand{\vk}{\vec{k}}

\newcommand{\vq}{\vec{q}}
\newcommand{\vp}{\vec{p}}

\def\D{\mathcal{D}}

\def\Tr{\mbox{Tr}}

\newcommand{\beqn}{\begin{eqnarray}}
\newcommand{\eeqn}{\end{eqnarray}}

\newcommand{\1}{\ensuremath 1\!\!1}
\renewcommand{\d}{\ensuremath\mathrm{d}}
\newcommand{\E}{\ensuremath\mathrm{e}}
\newcommand{\eps}{\ensuremath\varepsilon}
\newcommand{\Exp}[1]{\ensuremath\left< #1 \right>}
\newcommand{\fdg}{\ensuremath{\; \Big| \;}}
\newcommand{\GG}{\ensuremath\mathcal{G}}
\newcommand{\I}{\ensuremath \mathrm{i}}
\newcommand{\Int}{\ensuremath \mathrm{int\,}}
\newcommand{\norm}[1]{\ensuremath\left| #1 \right|}
\newcommand{\NN}{\ensuremath\mathbb{N}}
\newcommand{\OO}{\ensuremath\mathcal{O}}
\newcommand{\RR}{\ensuremath\mathbb{R}}

\newcommand{\LL}{\mathcal{L}}
\newcommand{\Matrix}[1]{\mathbf{#1}}
\newcommand{\SI}[1]{\hat{#1}}
\newcommand{\V}[1]{\boldsymbol{#1}}

\newcommand{\beqa}{\begin{eqnarray}}
\newcommand{\eeqa}{\end{eqnarray}}
\newcommand{\bp}{{\bf p}}
\newcommand{\bk}{{\bf k}}
\newcommand{\bq}{{\bf q}}
\newcommand{\bA}{{\bf A}}
\newcommand{\p}{\partial}
\newcommand{\too}{\rightarrow}

\newcommand{\Ord}{\mathcal{O}}
\newcommand{\reftitle}[1]{}

\begin{document}

\title{Panel discussion:  What {\it don't} we know about confinement?}


\author{R. Alkofer}
{
address={Institut f\"ur Physik, Universit\"at Graz, Universit\"atsplatz 5, 
  8010 Graz, Austria}
 }
\author{D. Diakonov}
{
address={Petersburg Nuclear Physics Institute, Gatchina 188300, St. Petersburg, Russia},
}
\author{J. Pawlowski}
{ address={Institut f\"ur Theoretische
    Physik, Universit\"at Heidelberg, Philosophenweg 16, 69120
    Heidelberg, Germany}
}

\author{H. Reinhardt}
{address={Universit\"at T\"ubingen, Institut f\"ur Theoretische Physik, Auf der Morgenstelle 14, 72076 T\"ubingen, Germany} }
\author{V. Zakharov}
{address={ ITEP, B. Cheremushkinskaya 25, Moscow, 117218 Russia} }
\author{D. Zwanziger}
{address={Physics Department, New York University, New York, NY 10003, USA } }
\author{(compiled by) J. Greensite}
{address={Physics and Astronomy Dept., San Francisco State Univ., San Francisco, CA 94132 USA } }

\begin{abstract}
The participants in this discussion session of the QCHS 9 meeting were each asked the following question:  ``What would be the most useful piece of information that you could obtain, by whatever means, that would advance your own program, and/or our general understanding of confinement?''  This proceedings contains a brief summary of each panel member's contribution to the discussion, provided by the panel members themselves. 
\end{abstract}

\maketitle


\section{Reinhard Alkofer}

\centerline{\it FUNCTIONAL APPROACHES TO CONFINEMENT}

\bigskip
 
\noindent{\bf On the question of the time-like quark-gluon coupling} \\

Jeff Greensite, when inviting to this round-table-dis\-cussion, raised the
question: What don't we know about Confinement? Although this question
might seem surprising it is very well-posed.

Let me follow for one moment the usually employed Gedankenexperiment of kicking
a quark out of a hadron. As most of you know such reasoning leads to the
de\-velopment of a picture where a flux tube or string is formed which pulls
either the quark back to the hadron or leads via string breaking and
de-excitation to the production of (many) hadrons. 

As clever as these arguments are presented, typically an aspect is overlooked
which, however, is only relevant in view of the available non-perturbative
techniques. Lattice calculations and functional approaches rest almost
exclusively on a formulation in Euclidean space-time. Processes which take
place at time-like momenta are only describable with some additional effort. 
Given the fact that practically all physical processes involve time-like
momenta this is of course a severe short-coming. {\it E.g.\/} taking your
prefered quark propagator with ``built-in confinement'' might work nicely when
considering static (or space-like) quantities but may fail miserably when
calculating production processes. (NB: A corresponding example can be found in
ref.\ \cite{Ahlig:2000qu}.)

An understanding of confinement can probably only be considered complete if it
includes knowledge about the coupling of gluons and quarks at time-like
momenta. One can only appreciate the infeasibility of the required task when
comparing to recent studies of the fully renormalized quark-gluon vertex
function at space-like momenta, see {\it e.g.\/}  ref.\ \cite{Alkofer:2008tt}
and references therein. First of all, one needs to notice that dynamical
breaking of chiral symmetry also changes the character of the quark-gluon
coupling: Not only vector-type chiral-symmetry preserving couplings are allowed
but also such type of couplings which are not invariant under chiral
transformations, especially scalar ones. Therefore, in order to discuss the
issue of Lorentz scalar confinement we have to consider confinement {\bf and}
dynamical breaking of chiral symmetry in one unified treatment. Mentioning
chiral symmetry one gets reminded that of course also the U$_A$(1) anomaly
might very well be  related to confinement, either  directly (see {\it e.g.\/}
\cite{Kogut:1973ab}) or indirectly due to the fact that some topologically
non-trivial field configurations are the ``confiners'' (see {\it e.g.\/}
\cite{Greensite:2003bk,DiGiacomo:2009gy,Diakonov:2009jq} and others). As the
anomaly reflects itself in the $\eta^\prime$ mass and in certain decays, again
we will only truly understand whether it is correlated with confinement when
we gain some insight into the coupling of quarks and glue at time-like momenta.

Therefore, either if we stay on purely theoretical playgrounds like
investigating strong-coupling phenomena or we are really getting serious in
calculating hadronic decays and/or production processes in terms of quarks and
glue: An insight into their time-like properties and their respective  mutual
coupling is required. This task certainly demands to generalize and enlarge the 
available ``tool box'' of non-perturbative  methods. \\

\noindent{\bf A remark on suitable gauges} \\

There are not so many non-perturbative tools available to treat Quantum Field
Theories at strong coupling, and hereby especially the strong interaction as
described by QCD. And there is basically only one method which does not require
gauge fixing: Lattice Monte-Carlo calculations. Corresponding calculations of
{\it e.g.\/} hadron spectra are certainly very useful but without refering to
the quark and glue substructure one is very restricted in learning about this
substructure. As quark and gluon correlation functions are only non-vanishing
if a gauge is fixed it can be very reasonable to fix a gauge.

Acknowledging the reasons for fixing a gauge (which is a pure necessity in
perturbation theory and functional approaches!) the following question arises: Are there
gauges which are more suitable than others? This question has already been
answered affirmatively  by Hugo Reinhardt in this discussion, and I want to
re-emphasize it again.

The Landau gauge has proven to be especially suited for the investigation of
Green functions, a guide to the literature can be {\it e.g.\/} inferred from the
reference list of \cite{Fischer:2008uz}. Even given the recent discussion on
different types of solutions (which has been also very present at this
Conference) one should not lose sight of how much knowledge has been
gained in the last decade.

Coulomb gauge is used in many investigations  (see the references in
\cite{Greensite:2010fd,Watson:2010fv}) and allowed for some fundamental insight 
\cite{Zwanziger:2002sh}. Therefore it is certainly worth investigating the
properties of quarks and gluons in this gauge, especially, as there is some
recent progress after Coulomb gauge QCD has proven in the past to be much more
complicated than expected, and this already at the perturbative level
\cite{Andrasi:2010dv}.

Last but not least, I would like to mention Maximally Abelian gauge, and
especially I like to point out that an infrared analysis of functional equations
has been performed recently \cite{Huber:2009wh}. And, in addition to lattice
calculations  in this gauge (see {\it e.g.\/} \cite{Mihari:2007zz} and
references therein), also the functional equations are a subject of current
investigations \cite{AWVMRA,Daniele}. This will allow further insight into the
conjecture of Abelian dominance and, in the best of all cases, into the
importance of chromomagnetic degrees of freedom as {\it e.g.\/}  monopoles and
center vortices. \\

\noindent {\bf A comment on the infrared behaviour of Landau gauge Green functions
and lattice calculations} \\

As I already mentioned the debate on the infrared behaviour of QCD Green
functions in the Landau gauge I would like to take the freedom to ask a question
which is probably only understandable to the experts. (Therefore all non-experts
might wish to skip this subsection.)

It is usually claimed that lattice calculations (in four space-time dimensions)
see only the so-called decoupling solution (sometimes also called the
``massive'' solution \cite{Aguilar:2008xm}) 
of functional equations and not the scaling solution. This is
certainly true if one takes into account only the available data for the gluon
propagator in the usual minimal Landau gauge.  On the other hand, going to the
extreme strong-coupling limit not only the infared exponents of the decoupling
but also the exponents of the scaling solution can be extracted 
\cite{Sternbeck:2008mv,Cucchieri:2009zt,Maas:2009ph}. To my opinion such an
extraction should be absolutely impossible if the scaling solution is definitely
absent and therefore the lattice calculations should not contain any trace of
it. But as both type of exponents are seen there is still something to be
understood here, especially in the view of the fact that there are dependencies
on the way the gluon field is discretized \cite{Sternbeck:2008mv}.

\bigskip

To summarize my contribution: Although the last years have seen tremendous
progress in our understanding of the infrared behaviour of QCD almost all of the
decisive questions cannot be definitely answered yet. However, if  one
acknowledges also our progress in developing new methods which are applicable
in the strong-coupling domain I am convinced that in the not-so-distant future
we will gain a basic understanding of the hardest problem of hadron physics, we
will eventually understand confinement.

\section{Dmitri Diakonov}

\centerline{\it HOW TO CHECK THAT DYONS ARE AT WORK?} 

\bigskip

There is multiple evidence from theory, lattice and phenomenology that dyons play
an important role in enacting confinement, and in inducing the deconfinement phase transition:
\begin{itemize}
\item In ${\cal N}=1$ supersymmetric theory it is precisely dyons that shape the vacuum, in particular the gluino condensate,
and it is an {\em exact} result there~\cite{Davies:1999uw,Davies:2000nw,Diakonov:2002qw}
\item Lattice studies show that zero fermion modes `jump' from one position in space to another as one varies
fermion boundary conditions -- precisely as one would expect from dyons carrying zero modes~\cite{Gattringer:2003uq}
\item a semiclassical picture of the vacuum populated by dyons gives an appealing explanation of all main features
associated with confinement: the area behavior of large Wilson loops, the asymptotic linear rising potential only for
nonzero $N$-ality probes, the cancelation of gluons in the free energy, and a 1$^{\rm st}$ order deconfinement
phase transition for all gauge groups except $SU(2)$, regardless of whether the group has a nontrivial center or not~\cite{Diakonov:2007nv,Diakonov:2010qg}.
\end{itemize}

In a few words, if the Polyakov line is, on the average, not an element of the group center, as in the confinement
phase, dyons appear as saddle points of the Yang--Mills (YM) partition function. Dyons are gluon field configurations
with asymptotic Coulomb-like chromo-electric and -magnetic fields. The ensemble of many dyons is similar to a
multi-component plasma. In particular, the dual (magnetic) gluons obtain a Debye mass. This is the physical reason
for the exponential decay of the Polyakov lines correlations, {\it i.e.} of the linear rising potential for heavy
probe `quarks'. The appearance of a mass for dual gluons is also responsible for the area behavior of large Wilson
loops. A surface spanning the loop is a source for a soliton of finite thickness, `made' of the dual fields.
The action per area of this soliton is the string tension which coincides, at low temperatures $T$,
with the string tension computed from the correlation function of the Polyakov lines~\cite{Diakonov:2007nv}.

The ensemble of dyons induce a nonperturbative energy that is a function of the Polyakov loop eigenvalues.
The function is such that its minimum, for any gauge group, is at a universal set of eigenvalues related
to the Weyl vector, the half-sum of positive roots of the Lie algebra for the gauge group~\cite{Diakonov:2010qg}.
For most gauge groups it implies that the trace of the Polyakov loop is zero in the lowest dimensional
representations (however there are subtleties related to the fluctuations about the minimum). This dyon-induced
nonperturbative potential energy competes with the well-known perturbative potential energy, also a function
of the Polyakov loop eigenvalues. The perturbative energy scales as $T^4$ with respect to the nonperturbative one.
Therefore, at some critical $T_c$ it prevails, and that is the mechanism for the deconfinement phase transition.
It happens irrespectively of what is the center of the gauge group. \\

There are presently several qualitative pictures of confinement being discussed. Apart from dyons, these
are Abelian monopoles and center vortices, see {\it e.g.}~\cite{Greensite:2009zz}. It may be that
all three pictures can be, in a sense, reconciled. Dyons may be the real physical objects that reveal
themselves as Abelian monopoles seen on the lattice in the Abelian gauges, such as the maximal Abelian gauge,
and also as center vortices observed in the center gauges, such as the maximal center gauge.

Indeed, in order to assemble many dyons together they need to have the same asymptotic field $A_4$ at spatial
infinity. That necessarily requires that dyons are in the `stringy' gauge where a singular Dirac string is
sticking out from each dyon. The Dirac strings are gauge artifacts: the action density is finite there. However,
they do carry a quantized Abelian magnetic flux. When, in a lattice simulation, one uses any variant of the
Abelian gauge, one identifies lattice magnetic monopoles as the sources of that flux. The exact position of
Abelian lattice monopoles varies somewhat as one varies gauge fixing -- in accordance with the fact that the
direction of the Dirac string sticking from a dyon is subject to a gauge choice. However, lattice magnetic monopoles
may well be a reflection of real physical objects, the dyons. It would be interesting to check it directly.

Furthermore, if one further restricts the gauge to a center gauge, center vortices are 
revealed~\cite{Greensite:2009zz}. They can be understood as the Dirac strings connecting dyons. 
There are gauges where a dyon has a Dirac string entering it, and another leaving it. 
See a recent study in Ref.~\cite{Bruckmann:2010bs} of the relation between dyons and vortices.

I should also mention a talk by Langfeld and Ilgenfritz at this
conference~\cite{Langfeld:2010ue}, who ``cooled'' 
lattice configurations keeping Polyakov loops fixed. Usually smearing the configurations by cooling 
kills confinement but in this study it is preserved. The interesting observation is that the 
``cooled'' configurations preserving confinement are mainly (anti)self-dual fields. 
Instantons and dyons are (anti)self-dual.

Concerning instantons, the quantum Coulomb interactions of dyons are such that they tend to glue up into
electric- and magnetic-neutral clusters which at low temperatures are hardly distinguishable from
instantons~\cite{Diakonov:2009jq}. The difference with the old `instanton liquid' model~\cite{Diakonov:1983hh}
is that ({\it i}) the Polyakov line is now nontrivial, ({\it ii}) the integration measure over collective
coordinates is invariant under permutation of dyons `belonging' to different instantons, and allows instantons
to overlap. These circumstances are critical for obtaining confinement that was absent in former old instanton
models.

Finally, I would like to point out a (simple) lattice measurement that may help to understand
the nature of the YM vacuum in general, and to demonstrate (or refute) the importance of dyons, in particular.
I suggest to measure the effective action for the gauge-invariant eigenvalues of the Polyakov line. The definition
is given in Eq.(3) of Ref.~\cite{Diakonov:2010qg}. In lattice setting, one puts all time links to be unity matrices
(the $A_4=0$ gauge), but make the spatial links periodic up to a gauge transformation with the matrix $L({\bf x})$
being the Polyakov line. Without loss of generality one can take it to be diagonal, as in Eq.(1) of
Ref.~\cite{Diakonov:2010qg}. One then simulates the ensemble of configurations with fixed `eigenphases'
$\mbox{\boldmath$\phi$}({\bf x})$ as boundary conditions. In particular, one can take $\mbox{\boldmath$\phi$}$
to be ${\bf x}$-independent. The partition function or the free energy itself is not calculable by Monte Carlo
methods but one can find the average plaquette and then integrate it over $\beta$ or temperature to obtain
the free energy. This will be the effective potential as function of $\mbox{\boldmath$\phi$}$. 
At large $T$ it is the well-known perturbative potential energy as function of $\mbox{\boldmath$\phi$}$. 
It is interesting to see how it looks like below and above $T_c$ for different groups. 

If, as we assume, dyons are of relevance, the minimum of the effective potential will be at a specific point
$\mbox{\boldmath$\phi$}$ proportional to the Weyl vector $\mbox{\boldmath$\rho$}$, for any gauge group,
see Eq.(5) of Ref.~\cite{Diakonov:2010qg}.  

It should be noted that if one just studies the distribution of the Polyakov line `eigenphases', it will be
in any case dominated by the Haar measure weight that governs the local ultraviolet quantum fluctuations. 
In order to see the smooth potential energy as function of $\mbox{\boldmath$\phi$}$, one really needs to consider
the case of a constant or slowly varying Polyakov lines using, for example, the setting described above.\\



\section{Jan PawlowskI}

\centerline{\it CONFINEMENT AND THE PHASE}
\centerline{\it DIAGRAM OF QCD}

\bigskip

An even partial answer to Jeff's question ``What don't we know about
confinement?'' certainly gives rise to new ones, as is the case for
all good questions. In this little comment I would rather like to
dwell on these new questions, as well as on weakened versions of the original
question.

In my opinion the question ``What don't we know about confinement?'' is
closely related to the question ``How do we detect confinement?''. In
Yang-Mills theory the latter question seems to be simple to answer:
watch out for the linear potential between static quark sources! Such
a linear potential can be extracted from e.g. the area law for Wilson
loops for large areas, or, at finite temperature, from the large
distance behavoir of the correlation function of a Polyakov loop and
an adjoint Polyakov loop, just to name two of the many
possibilities. This is fine as long as one does not add the question
of the confinement mechanism. I think it is precisely here where the
problems start. The above-mentioned operators are inherently
non-local, and all attempts to describe the confinement mechanism are
based on at least semi-local degrees of freedom, mostly effective
theories of topological defects. Loosely speaking one could say that
the more local the description gets the less convincing it is. In any
case a quasi-local description implies a specific parametrization of
the theory (a choice of a coordinate system), at least if quantum
fluctuations are taken into account. Evidently such a choice is
closely related to a gauge fixing procedure, and it is a hard-learned
lesson in the past that the quest for the mechanism of confinement
goes hand in hand with explicit or implicit gauge fixings as well as
the question of their interrelation.

This leaves us with the question of how to detect confinement, in
particular in formulations of QCD which are explicitly
gauge-fixed. Again there have been a plethora of suggestions, for
example based on the construction of the heavy quark potential, or on
signatures of the absence of colored states in the physical Hilbert
space. More recently much progress has also been made in directly
constructing observables that are sensitive to the confining
properties of QCD. 

Still, in my opinion the above question is not fully clarified yet,
and this fact is hampering the progress in our understanding of
confinement in QCD with dynamical quarks. Due to string breaking the
above definitions do not work anymore, and a clear signature of
confinement is missing. This problem gets even more severe if we
switch-on chemical potential/finite density. Then the question would
be first ``What does confinement mean in a dense medium of
quarks/hadrons?''  and then followed by ``How do we detect it?''.  Again
these apparently simple questions are very difficult to answer, in
particular on the basis of the other unresolved questions posed above,
starting with Jeff's original one. Of course one is attempted to
discard the latter questions as superfluous as we expect in this case
the matter properties of QCD to dominate the physics. However, there
seems to be a rather tight relation between chiral symmetry breaking
and confinement (however we detect it) that even persists at finite density. In conclusion we
will not understand this relation if we do not learn of how to answer
the above questions about the glue sector in detail.

Let me finish this little comment with toning down Jeff's original
question to ``What do we {\it want} to know about confinement?'' and
by collecting a couple of the answers given here: 

\begin{itemize}
\item What is the confinement mechanism? 
\item How do we detect confinement?
\item How does confinement relate to chiral symmetry breaking?
\item What does confinement mean in a dense medium of quarks/hadrons?
\end{itemize}

\section{Hugo Reinhardt} 

\centerline{\it CONFINEMENT $-$ WHAT WE DO KNOW,}
\centerline{\it AND WHAT WE DO NOT KNOW}

\bigskip

Confinement is one of the millenium problems formulated by the Clay Mathematical Institute. To receive the corresponding prize requires to prove besides confinement, i.e. the absence of quarks and gluons from the physical spectrum, the existence of a mass gap and spontaneous breaking of chiral symmetry. In the physics community the generally accepted confinement criterium is that the temporal Wilson loop (order parameter) develops an area law
\beq
\label{179-1}
\langle W (C) \rangle = \left\langle \mathrm{tr} \exp \bigg( - \oint\limits_C A \bigg) \right\rangle \sim \exp \lk - \sigma \Sigma (C) \rk 
\eeq
or equivalently the spatial 't Hooft loop \cite{'tHooft:1977hy} (disorder parameter) with continuum representation \cite{Reinhardt:2002mb}
\beq
\label{G2}
\langle V (C) \rangle = \left\langle \exp \lk - \int \vec{\cA} (C) \vec{\Pi} \rk \right\rangle \sim \exp \lk - \kappa L (C) \rk
\eeq
 develops a perimeter law. Here, $\Pi (x) = \delta / i \delta A (x)$ is the momentum operator of the gauge field and $\cA (C)$ is the gauge field of a center vortex whose magnetic flux is located at the loop $C$.

\bigskip
\noindent {\bf What we do know} \\

The 't Hooft loop operator in eq. (\ref{G2}) is a center vortex generator, which indicates that confinement must be related to the center of the gauge group. Indeed, from lattice calculations we know that in an $SU (N)$ gauge theory an asymptotic string tension is obtained in all representations with odd $N-$ality, i.e. which transform non-trivially under the center of the gauge group \cite{Greensite:2003bk}. Though we do not yet fully understand confinement, several pictures of confinement have emerged, which are supported by lattice results, see table \ref{Tabelle}.

The first two pictures rely on the condensation of magnetic monopoles and center vortices, respectively, while the Gribov (-Zwanziger) picture requires an infrared diverging ghost form factor $d$ defined by 
\beq
\label{199-3}
\langle (- D \partial)^{- 1} \rangle = d / (- \Delta) \hk .
\eeq
Here $- D \partial$ is the Faddeev-Popov operator in Landau or Coulomb gauge with $D$ being the covariant derivative in the adjoint representation.

Empirically it was found on the lattice that the different pictures of confinement are linked to each other: Magnetic monopoles are located on center vortices, \cite{Greensite:2003bk}, and both configurations are on the Gribov horizon in both Coulomb and Landau gauge \cite{Greensite:2004ke}. Removing center vortices removes the magnetic monopoles and makes the ghost form factor infrared finite \cite{Gattnar:2004bf}. Furthermore, in the absence of magnetic monopoles, center vortices are topologically trivial in the sense that they have vanishing Pontryagin index \cite{Reinhardt:2001kf}. Topologically trivial field configurations cannot account for the topological susceptibility and spontaneous breaking of chiral symmetry. 

We have lattice evidence that chiral symmetry breaking is lost when the ``confiners'' like center vortices and magnetic monopoles are removed from the Yang-Mills ensemble \cite{Greensite:2003bk}. By the Banks-Casher relation \cite{Banks:1979yr} a non-vanishing quark condensate (the order parameter of chiral symmetry breaking) requires a non-zero quark level density at zero virtuality. As shown in ref. \cite{Gattnar:2004gx}  the quark spectrum develops a gap at zero virtuality when center vortices are removed from the Yang-Mills ensemble.

We also know that confinement is realized only in the low temperature and low matter density phase and a deconfinement phase transition occurs when the temperature or matter density exceeds certain critical values. Furthermore, in the center vortex picture (and similarly in the monopole picture) the deconfinement phase transition is a depercolation transition \cite{Engelhardt:1999fd}.

\begin{table}
\label{Tabelle}
\begin{tabular}{c|c|c}
scenario & dom. field config. & gauge\\
\hline
dual supercond. & mag. monopoles & Max. Abelian \\
vortex cond. & center vortices & Max. Center \\
Gribov- & Gribov horizon & Landau/Coulomb \\
Zwanziger &&
\end{tabular}
\caption{Survey of confinement scenarios}
\end{table}

\bigskip
\noindent {\bf What we do not know} \\

So far, we do not have a rigorous proof of confinement in Yang-Mills theory in the sense of the Clay Mathematical Institute. Furthermore, we have no gauge invariant confinement picture. 
What is common to all the confinement pictures listed in table \ref{Tabelle} is that they rely on specific gauges. Of course, confinement is a physical phenomenon, which does not depend on a specific gauge choice.  However, the confinement mechanism may be easier to reveal in one gauge than in another. There is nothing wrong in choosing a convenient gauge (choosing a gauge is nothing more than choosing specific ``coordinates'') but one has to show that the confinement picture found in one gauge persist when the gauge is changed. 
The ultimate confinement scenario should be formulated in a gauge invariant way. It is clear that the Gribov-Zwanziger scenario intrinsically requires a gauge fixing. Furthermore, the Gribov-Zwanziger mechanism seems to be realized in Coulomb gauge but not in Landau gauge. 
We also do not know the nature of the deconfinement phase transition in Gribov's confinement scenario. Clearly, the Gribov horizon of Coulomb gauge is independent of the temperature.

Finally, we do not know yet how spontaneous breaking of chiral symmetry is accomplished in the various pictures of confinement. If confinement and chiral symmetry breaking are triggered by the same mechanism then the restoration of chiral symmetry should precisely occur at the deconfinement phase transition. This is confirmed on the lattice \cite{Bonati:2009zz}, \cite{Bornyakov:2009qh}. \\

\noindent { \bf Results of the Hamilton approach to Yang-Mills theory in Coulomb gauge} \\

During the last years our group has worked on Yang-Mills in Coulomb gauge, both in the continuum and on the lattice and the obtained results strongly support Gribov's scenario of confinement. In the continuum an approximate variational solution of the Yang-Mills Schr\"odinger equation was carried out with the following results: 
\begin{enumerate}
\item
The gluon energy $\omega (k)$ is infrared divergent and can be nicely fitted by Gribov's formula 
\beq
\label{232-4}
\omega (k) = \sqrt{\vec{k}^2 + M^4 / \vec{k}^2} \hk ,
\eeq
where $M \approx 860 MeV$ \cite{Gribov:1977wm}, \cite{Feuchter:2004mk}.
The same result is also obtained on the lattice \cite{Burgio:2008jr}. Hence, gluons are absent from the physical spectrum, which is a manifestation of gluon confinement.

\item The 't Hooft loop is found to obey a perimeter law \cite{Reinhardt:2007wh}.

\item A linear rising non-Abelian Coulomb potential was found \cite{Epple:2006hv}, which is necessary but not sufficient for the area law in the Wilson loop.

\item The inverse ghost form factor $d^{- 1} (k)$ was shown to represent the dielectric function of the Yang-Mills vacuum \cite{Reinhardt:2008ek}. As a consequence, the horizon condition $d^{- 1} (k = 0) = 0$ implies that the Yang-Mills vacuum is a dual superconductor, which relates Gribov's confinement scenario with the monopole condensation picture.

\end{enumerate}

The results of the variational approach were not only confirmed by lattice calculations but also by alternative approaches to continuum Yang-Mills theory like Dyson-Schwinger equations \cite{Watson:2010cn} and renormalization group flow equations \cite{Leder:2010ji}.

\section{Valentin Zakharov}

\centerline{\it SCALAR FIELDS AND CONFINEMENT}

\bigskip

 Scalar, or Higgs fields are an indispensable part of the Standard Model.
 Another scalar field, the inflaton, is a common device used to explain cosmological inflation.
 The theory of confinement is in the same sequence: it starts with an analogy
 to superconductivity and introduces a scalar field
 $\phi_{GL}$ a la (relativistic version of)
 Ginzburg-Landau theory.
 In particular, for the tension of the string connecting heavy quarks
 one would expect:
 \begin{equation}\label{expectation}
 \sigma_{string}~\sim~<\phi_{GL}>^2~,
 \end{equation}
 where $\phi$ is a scalar field.

In each case we deal with field theories which are, at first sight, well defined.
 And nevertheless it is common knowledge that in fact such theories are inconsistent because
 of the quadratic divergences. Thus, we have to expect the field theories to be replaced by something
 else.

It is only in case of confinement, however,  that there exists an independent
 source of information on the nature of the scalar field, that is lattice simulations.
 And, indeed, the interpretation
 of the data (Chernodub and Zakharov, 2003) looks unusual:
 \begin{eqnarray}\label{remarkable}
 <|\varphi_{magn}|^2>~\sim~\Lambda_{QCD}^2, \\\nonumber
 (<\varphi_{magn}>)^2~\sim~\Lambda_{QCD}^{3}a~,
 \end{eqnarray}
where $\varphi_{magn}$ is the magnetically charged scalar responsible for
confinement, as seen on the lattice,
 $a$ is the lattice spacing, $a\to 0$ in the continuum limit.
 Thus, in contrast to (\ref{expectation}), the vacuum expectation value
 vanishes in fact and, nevertheless, suffices to ensure confinement at any
 $a\neq 0$!

 From dimensional considerations alone it is obvious that
 the Eqs (\ref{remarkable}) are indicating to strings, or 2d field theory.
 Moreover, it is also seen in the simulations that the scalar-particle
 trajectories cover densely 2d surfaces, or strings (defined independently).
 Thus, the UV quadratic divergence seems to be avoided by invoking strings.
 (It is remarkable that
 it is known since long (Atick and Witten, 1988) that strings
 are much 'softer' in terms of effective degrees of freedom than field
 theories. In that case,  two powers
 of temperature were missing in the limit of high temperatures.
 At $T=0$, two powers of the UV cut off are absent from $<|\phi|^2>$.)

 In the continuum language, the most recent candidate for a 'soft' scalar field
 is the Horava-Lifshitz scalar, (Horava, 2009). It is assumed that at {\it short distances}
 theory becomes time-space asymmetric, $ L\sim (\partial_t\phi)^2-(\Delta\phi)^2$.
 Furthermore, one   argues that the effective field theory is in fact two-dimensional at short distances.
In this sense, the solution for the scalar is similar to what we have just discussed
with reference to the lattice.

The lattice data refer to the Euclidean case and there can be no time-space asymmetry.
While the Horava-Lifshitz scalar refers to the Minkowski space. The matching can be discussed
only in terms of matrix elements, like $<|\phi|^2>$, which can be safely continued
to Minkowski space. If the matching observed is meaningful indeed, then there arises
an intriguing possibility that confinement effects in Minkowski space and at short distances do not observe Lorentz invariance, whatever it might mean.
 
\newpage

\section{Dan Zwanziger}

\centerline{\it CAN THE FREE ENERGY WITH SOURCES}
\centerline{\it BE MEASURED ON THE LATTICE?}

\bigskip

	Jeff Greensite has posed to the panelists the question ``From your own perspective, what would be the most useful piece of information that you could obtain (from either lattice experiments, real experiments, analytical methods, or maybe from God) that would advance your own program, and/or our general understanding of confinement?''  I will suggest a lattice measurement of a perhaps novel type, and explain why it appears useful.  Before doing so, I frankly confess that I don't know whether the suggested measurement is feasible in practice.    

	Recent numerical studies on large lattices of the gluon propagator in momentum space $D(k)$, reviewed recently in \cite{Cucchieri:2010.03}, yield finite values for $D(0)$, in apparent disagreement with the theoretical expectation that $D(0) = 0$, originally obtained by Gribov \cite{Gribov:1977wm}, and argued in \cite{Zwanziger:1991}.  Upon reviewing the argument \cite{Zwanziger:1991} which leads to this result, one hypothesis stands out which should perhaps be dropped in view of the apparent disagreement.  This is the hypothesis that the free energy $W(J)$ in the presence of sources $J$ is analytic in $J$.  This is an important point because a non-analyticity in the free energy is characteristic of a change of phase.  

	The free energy $W(J)$ enters the picture because it is the generating function of the connected gluon correlators.  In particular the gluon propagator $D_{x,y}$, is a second derivative of $W(J)$ at $J = 0$,
\beq
D_{x, y} = {\p^2 W(J) \over \p J_x \p J_y}\Big|_{J=0}.
\eeq
Here a condensed index notation is used, where $J_x$ represents $J_\mu^b(x)$, $\mu$ is a Lorentz index, and $b$ is a color index, and we write
\beq
(J, A) = \sum_x J_x A_x =  \int d^Dx \ J_\mu^b(x) A_\mu^b(x).
\eeq
The free energy $W(J)$ in the presence of sources $J$ is given by
\beqa
\exp W(J) & = & \langle \exp(J, A) \rangle
\nonumber \\
& = & \int_\Omega dA \ \rho(A) \exp(J, A).
\eeqa
For simplicity we have written this in continuum theory, but we have in mind a numerical study on the lattice, and the lattice analog of each formula should be obvious.  The integral over $A$ is effected in Landau gauge $\p_\mu A_\mu = 0$.  The domain of integration is restricted to the Gribov region $\Omega$, a region in $A$-space where the Faddeev-Popov operator is non-negative, $M(A) \equiv - \p_\mu D_\mu(A) \geq 0$.  In practice, the positive, normalized, Euclidean probability density $\rho(A)$ will depend on the numerical algorithm which is used for gauge fixing.  

	[To give a concrete example, one may imagine an idealized gauge fixing -- not attainable in practice -- to the absolute minimum of the minimizing functional $F_A(g) = || {^g}A||^2$ on each gauge orbit.  In this case, in the absence of quarks, $W(J)$ is given by
\beq
\exp W(J) = \int_\Lambda dA \ \rho(A) \exp(J, A),
\eeq    
where the fundamental modular region $\Lambda \subset \Omega$ is a subset of the Gribov region, $\rho(A)$ is given by the Faddeev-Popov formula, 
\beq
\rho(A) = N \delta(\p \cdot A) \det(M(A)) \exp[ - S_{YM}(A)], 
\eeq
and $S_{YM}(A)$ is the Yang-Mills action.]

	The gluon $n$-point function is the n-th derivative of $W(J)$ at $J = 0$.  To elucidate the analytic structure of $W(J)$ it would be useful to measure $W(J)$ itself for finite values of $J$, instead of only of few of its derivatives at $J = 0$.  For example, one could take
\beq
J_\mu^b(x) = H_\mu^b \cos(k \cdot x),
\eeq
and measure
\beq
\exp W(k, H) =
\langle \ \exp[ \sum_x H_\mu^b \cos(k \cdot x) A_\mu^b(x) ] \ \rangle,
\eeq
for a set of values of the parameters $k_\mu$ and $H_\mu^b$.  By analogy with spin models, $H_\mu^b$ may be interpreted as the strength of an external magnetic field, with a color index $b$, which is modulated by a plane wave $\cos(k \cdot x)$.  The measurement is done using the standard gauge-invariant Wilson action followed by gauge fixing to the Landau gauge by a minimization algorithm.  

	In \cite{Zwanziger:1991} it was shown that the free energy per unit (Euclidean) volume satisfies the bound
\beq
{W(k, H) \over V} \leq c |k| |H|,
\eeq  
where $c$ is a certain constant.  It would be of interest to investigate numerically whether this bound is saturated asymptotically at low $k$.  If so, it would be a remarkable non-analyticity because the bound is linear in $|H|$, whereas normally one expects $W(k, H)$ to be a power series in $H$, with leading term of order $H^2$.  However such analytic behavior leads to the above-mentioned disagreement with the lattice data.

	It may be that importance sampling based on $\rho(A)$ gives poor results for $\langle \exp(J, A) \rangle$ because of the exponential character of the observable $\exp(J, A)$, although it is in fact real and positive.  If so, one could alternatively try to measure $A(x)$ in the presence of finite source $J$,
\beq
\label{average}
\langle A(x) \rangle|_J = { \int_\Lambda dA \ \rho(A) \exp(J, A) \ A(x) \over \int_\Lambda dA \ \rho(A) \exp(J, A)},
\eeq
using importance sampling with the positive weight
\beq
\rho_J(A) = \rho(A) \exp(J, A).
\eeq	
This has the difficulty that the importance sampling must be done after gauge fixing, using the Faddeev-Popov weight
\beq
\rho(A) = \delta(\p \cdot A) \det(M(A)) \exp[ - S_{\rm YM}(A)].
\eeq


\bibliographystyle{aipprocl} 



\end{document}